\documentclass[aps,
prl,
reprint,
numerical,
superscriptaddress
]{revtex4-1}

\usepackage{amsmath}
\usepackage{color}
\usepackage{graphicx}
\usepackage{natbib}
\usepackage{hyperref}

\usepackage{siunitx}
\usepackage{changes}

\begin{document}

\title{Coupling thermal atomic vapor to an integrated ring resonator} 

\author{Ralf Ritter}
\thanks{R. Ritter and N. Gruhler contributed equally to this work.}
\affiliation{5. Physikalisches Institut and Center for Integrated Quantum Science and Technology, Universit\"at Stuttgart, Pfaffenwaldring 57, 70550 Stuttgart, Germany}
\author{Nico Gruhler}
\thanks{R. Ritter and N. Gruhler contributed equally to this work.}
\affiliation{Institute of Nanotechnology, Karlsruhe Institute of Technology, 76344 Eggenstein-Leopoldshafen, Germany}
\affiliation{Institute of Physics, University of M\"unster, M\"unster 48149, Germany}
\author{Wolfram Pernice}
\affiliation{Institute of Nanotechnology, Karlsruhe Institute of Technology, 76344 Eggenstein-Leopoldshafen, Germany}
\affiliation{Institute of Physics, University of M\"unster, M\"unster 48149, Germany}
\author{Harald K\"ubler}
\affiliation{5. Physikalisches Institut and Center for Integrated Quantum Science and Technology, Universit\"at Stuttgart, Pfaffenwaldring 57, 70550 Stuttgart, Germany}
\author{Tilman Pfau}
\affiliation{5. Physikalisches Institut and Center for Integrated Quantum Science and Technology, Universit\"at Stuttgart, Pfaffenwaldring 57, 70550 Stuttgart, Germany}
\author{Robert L\"ow}

\email{r.loew@physik.uni-stuttgart.de}

\affiliation{5. Physikalisches Institut and Center for Integrated Quantum Science and Technology, Universit\"at Stuttgart, Pfaffenwaldring 57, 70550 Stuttgart, Germany}

\date{\today}

\begin{abstract}
Strongly interacting atom-cavity systems within a network with many nodes constitute a possible realization for a quantum internet which allows for quantum communication and computation on the same platform. To implement such large-scale quantum networks, nanophotonic resonators are promising candidates because they can be scalably fabricated and interconnected with waveguides and optical fibers. By integrating arrays of ring resonators into a vapor cell we show that thermal rubidium atoms above room temperature can be coupled to photonic cavities as building blocks for chip-scale hybrid circuits. Although strong coupling is not yet achieved in this first realization, our approach provides a key step towards miniaturization and scalability of atom-cavity systems.    
\end{abstract}

%
\maketitle 

Light matter interaction in a cavity is one of the paradigms of quantum optics. Coupling a two level atom to a resonator mode has established the field of cavity quantum electrodynamics (cQED) \cite{Kimble1998,Haroche1993,Haroche2013}. Pioneering cavity QED experiments in the optical domain used thermal atomic beams to detect optical bistability as well as non-classical correlation functions in the transmitted light field of a high finesse cavity \cite{Rempe1991}. Coupling atoms to microresonators was demonstrated early-on with dilute cesium vapor at room temperature through the interaction with  whispering gallery modes of a fused silica microsphere \cite{Vernooy1998}. The strong coupling regime with integrated devices has been explored by interfacing cold atoms with nanophotonic resonators \cite{Aoki2006,Lukin2014} and photonic crystal waveguides \cite{Goban2014}.  While progress on the miniaturization of cold atom experiments has been reported \cite{Anderson2004}, their scaling in combination with cavity networks remains challenging. In contrast, thermal vapor cells allow for a scalable approach to quantum networks when combined with resonant nanophotonic circuits.

Thermal atoms have been coupled to guided light modes, e.g. in integrated hollow waveguides \cite{Schmidt2007,Schmidt2010}, hollow core fibers \cite{Epple2014,Gaeta2010}, tapered nano-fibers \cite{Rauschenbeutel2011,Franson2010,Shariar2008}, as well as solid core waveguides \cite{Levy2013,Ritter2015}, and are envisaged to serve as building blocks for a combined atom-nanophotonic network.   
Besides solid state physical systems like quantum dots, defect centers in crystals or single molecules embedded in host matrices, atoms provide a uniquely narrow distribution of transition frequencies. Therefore they are well suited for realizing quantum networks when coupled to optical cavities. Here we demonstrate the interaction of thermal atoms with ring resonators integrated into nanophotonic circuits, which has been proposed theoretically recently \cite{Stern2012}.  

Although the coupling between atoms and the cavity mode in our first experiments is still much lower compared to cavity QED experiments, this demonstration solves two major technical problems: first, we increase integration density and design flexibility by coupling in and out of a waveguide with compact Bragg couplers, which can be placed anywhere on the chip. Second, we achieve protection of the waveguide material (silicon nitride (Si$_3$N$_4$)) against chemical deterioration by the aggressive alkali atoms (here rubidium) and thus enable long-term usage of the atom-clad platform. We estimate that these advances allow for the design of robust atom-cavity networks which have the potential to reach cooperativity factors larger than one.  

\begin{figure*}
	\centering
    \includegraphics[scale=1]{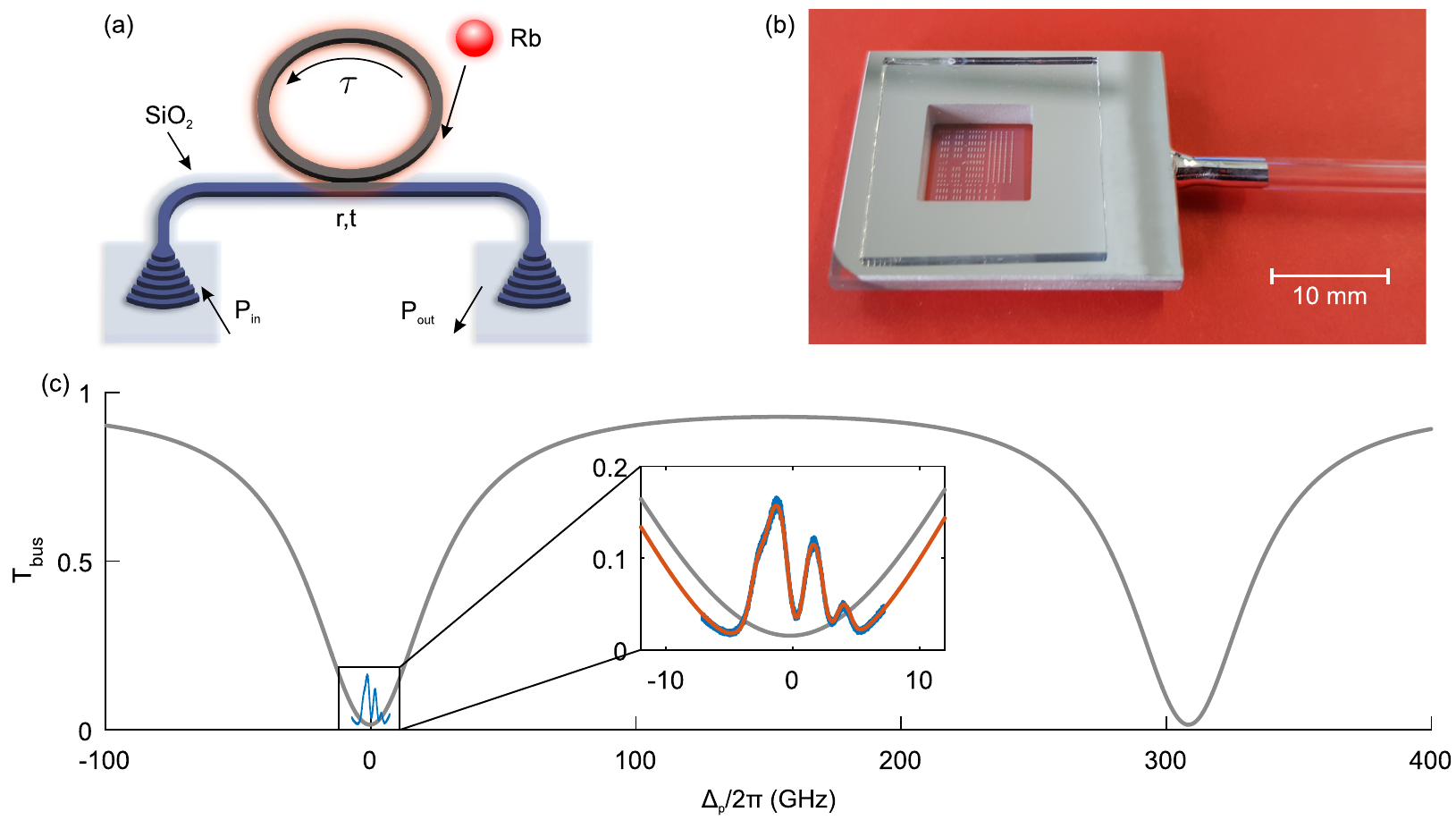}
    \caption{Coupling an integrated ring resonator to atomic vapor. (a) Schematic diagram of the ring resonator surrounded by rubidium vapor. The probe light $P_{\textrm{in}}$ is injected into the bus waveguide via a Bragg coupler and subsequently coupled to the ring with coupling parameters $r$ and $t$. The output $P_{\textrm{out}}$ is detected with a photo multiplier tube (PMT). (b) Photograph of the vapor cell and the bonded optical chip with the rubidium reservoir on the right. (c) The gray curve shows the calculated bus waveguide transmission without contribution of the atoms. The transmission is modified in presence of the atoms (blue curve), which is shown more detailed with a fit of the model (red curve) in the inset.}
    \label{fig:fig1}
\end{figure*}  

\begin{figure}[b!]
\centering
    \includegraphics[scale=1]{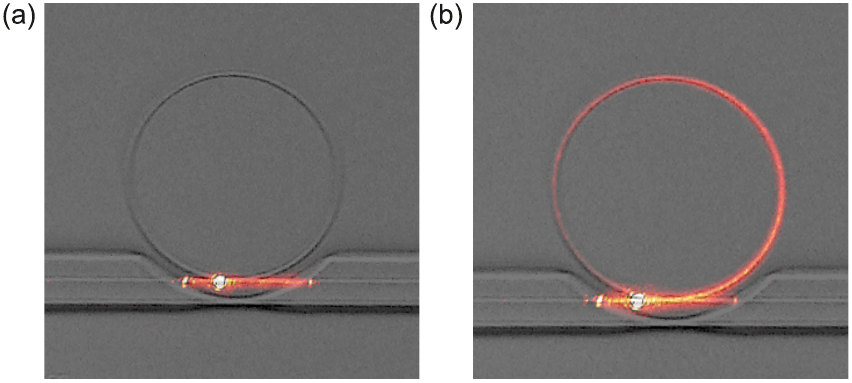}
    \caption{Photographs of the ring for an off-resonant situation (a) and for $\omega_R\approx\omega_0$ (b).The interaction of the evanescent field with the rubidium vapor is directly visible due to the fluorescence light of the atoms, which is color coded in these pictures.}
    \label{fig:fig2}
\end{figure}

For the experiments described in this work, we used a Si$_3$N$_4$ ring resonator with a radius of \SI{80}{\micro\meter} on a borosilicate substrate. As illustrated in Fig.\,\ref{fig:fig1}a, the ring is excited and probed via a bus waveguide terminated with grating couplers for in- and out-coupling of light. To restrict the atom light interaction mainly to the ring resonator, all remaining parts are completely covered with a \SI{600}{\nano\meter} thick layer of silicon dioxide (SiO$_2$), except for the short coupling region between the bus waveguide and the ring. As rubidium atoms sticking to the waveguide surface increase transmission losses \cite{Ritter2015} and therefore cause the resonances of the ring to disappear, we additionally cover the structures with a \SI{9}{\nano\meter} thick sapphire (Al$_2$O$_3$) coating by means of atomic layer deposition. With this protection coating the resonances remain visible, although their linewidth is still increased after rubidium exposure. This way the devices are still usable after several months without showing further degradation of the optical performance, which is essential for realizing more advanced circuits. In the Supplementary Information we present details of the fabrication procedure and the transmission properties of the devices.

A nanophotonic chip containing multiple individual rings is integrated into a rubidium vapor cell using triple stack anodic bonding \cite{Daschner2014} as shown in Fig.\,\ref{fig:fig1}b. A rubidium reservoir is attached to the cell and can be heated independently from the chip to control the atom density. This gives a small and very convenient system independent of large apparatus in contrast to cold atom experiments. By varying the temperature of the chip, the ring resonance frequency $\omega_{\text{R}}$ can be tuned with respect to the atomic resonance $\omega_0$ (center of mass frequency of the rubidium D$_2$ line). In order to approach critical coupling, where internal resonator loss and coupling loss are equal, we select a device with an appropriate distance between the bus waveguide and the ring such that near zero transmission on resonance is achieved.

\begin{figure*}[t!]
\centering
    \includegraphics[scale=1]{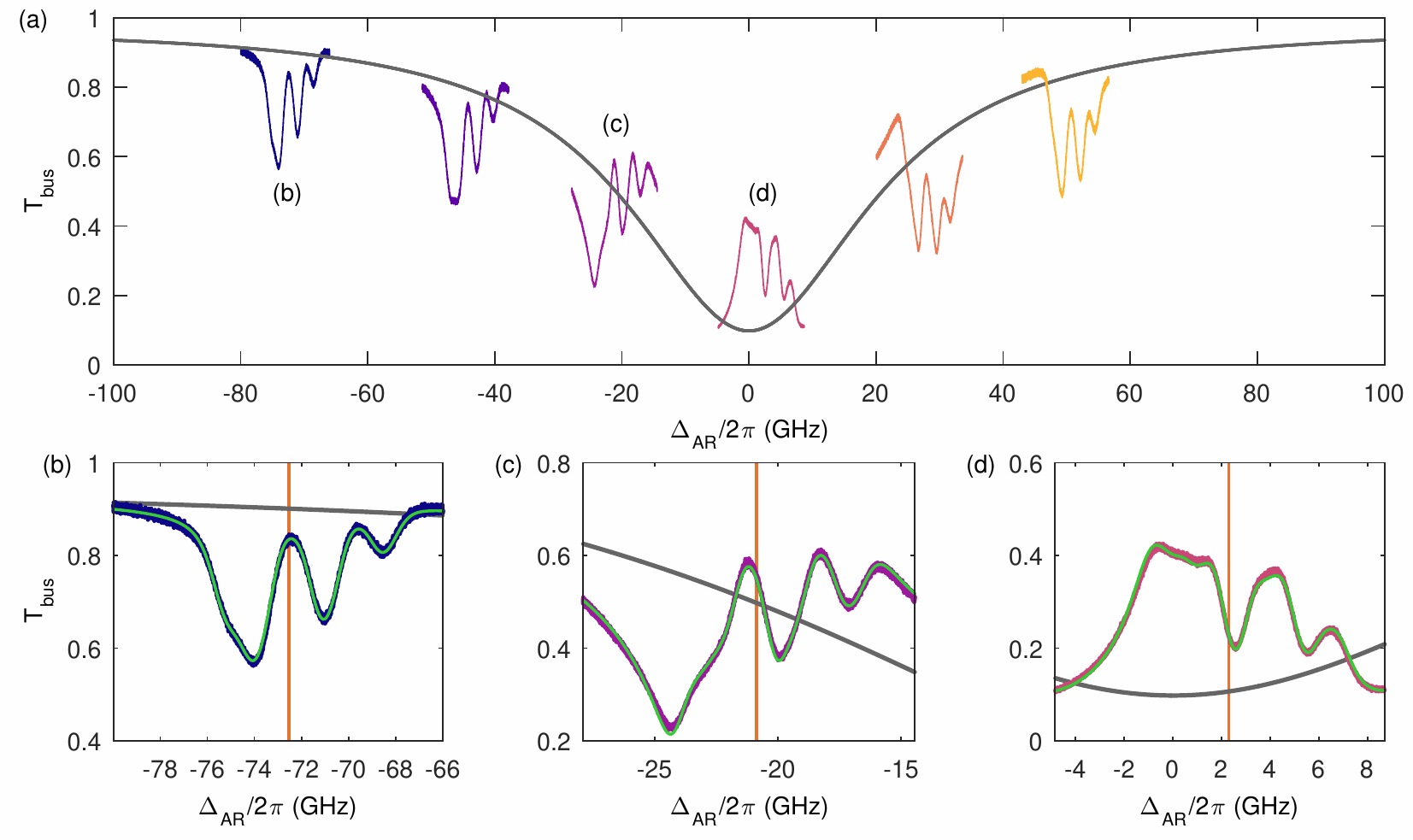}
    \caption{Transmission measurements for various atom-resonator detunings $\Delta_{\textrm{AR}}$. (a) The gray curve shows the bus waveguide transmission without atomic contribution, as calculated from fit parameters. The superimposed spectra show the transmission for an atomic density of $n_g\approx\SI{e14}{\centi\meter^{-3}}$ at corresponding $\Delta_{\textrm{AR}}$. (b-d) Individual transmission spectra for an off-resonant ring (b), on the slope of the resonance (c) and close to resonance $\omega_R\approx\omega_0$ (d) with fits of the model (green curves). The vertical lines indicate the positions of the center of mass frequency of the rubidium D$_2$ line, $\omega_0$.} 
    \label{fig:fig3}
\end{figure*} 

The inset of Fig.\,\ref{fig:fig1}c shows a typical transmission spectrum of the bus waveguide at a moderate atom density ($n_g\approx\SI{2e13}{\centi\meter^{-3}}$), where the ring resonance is centered to the atomic resonance and the probe frequency $\omega_{\textrm{p}}$ is scanned over the rubidium D$_2$ line with the detuning $\Delta_{\textrm{p}} = \omega_0-\omega_{\textrm{p}}$. The transmission of the device without contribution of the atoms is calculated from fit parameters of our model (Supplementary Information) and displayed in the background of Fig.\,\ref{fig:fig1}c over more than one free spectral range ($\textrm{FSR}/2\pi\approx \SI{308}{\giga\hertz}$) to visualize the bandwidth proportions. Owing to the motion of the atoms, the spectral line shape is broadened due to the Doppler effect and the short transit time of the atoms traveling through the evanescent field. Both absorptive and dispersive properties of the atoms play a role, when they interact with the resonator mode. The absorption of the light field lifts the critical coupling condition, leading to an increase in transmission for $\omega_R\approx\omega_0$, whereas the real part alters the round trip phase shift, leading to a shift of the ring resonance to lower (higher) frequencies on the red (blue) side of the atomic resonance. Note that the total signal is always a combination of the ring signal and the absorption signal from the \SI{100}{\micro\meter} long uncovered part of the bus waveguide. The interaction of the atoms with the ring does not only manifest itself in the transmission signal, but is also directly visible in the fluorescence of the atoms as shown in Fig.\,\ref{fig:fig2}a and b for an off-resonant and a resonant situation, respectively. 

In order to investigate the transmission behavior at different positions within the ring resonance, we performed a series of measurements where we thermally tune the ring resonance frequency to several values of the atom-resonator detuning $\Delta_{\textrm{AR}}=\omega_0-\omega_{\textrm{R}}$, while scanning the probe laser over the rubidium D$_2$ line. Figure\,\ref{fig:fig3} presents the results of these measurements for an atom density of $n_g\approx\SI{e14}{\centi\meter^{-3}}$. In Fig.\,\ref{fig:fig3}a the bus waveguide transmission spectra are placed at the corresponding positions of the ring resonance feature, as determined from fits to the data. Figure\,\ref{fig:fig3}b-d show selected transmission data for three values of $\Delta_{\textrm{AR}}$ together with their respective fitting curves. In the off-resonant case shown in Fig.\,\ref{fig:fig3}b the transmission spectrum is dominated by absorption in the uncovered part of the bus waveguide, since there is almost no coupling of the probe into the ring, which is also clearly visible in Fig.\,\ref{fig:fig2}a.

At the slope of the resonance (see Fig.\,\ref{fig:fig3}c), the signal reveals the dispersive nature of the atoms, since a small change in the real part of the atomic susceptibility leads to a large modulation of the ring transmission. The third characteristic feature is found on resonance ($\omega_R\approx\omega_0$) where the additional losses induced by the atomic absorption lead to increased transmission. This situation is shown in Fig.\,\ref{fig:fig3}d, where the transmission enhancement amounts approximately $40\%$. 

\begin{figure}
\centering
    \includegraphics[scale=1]{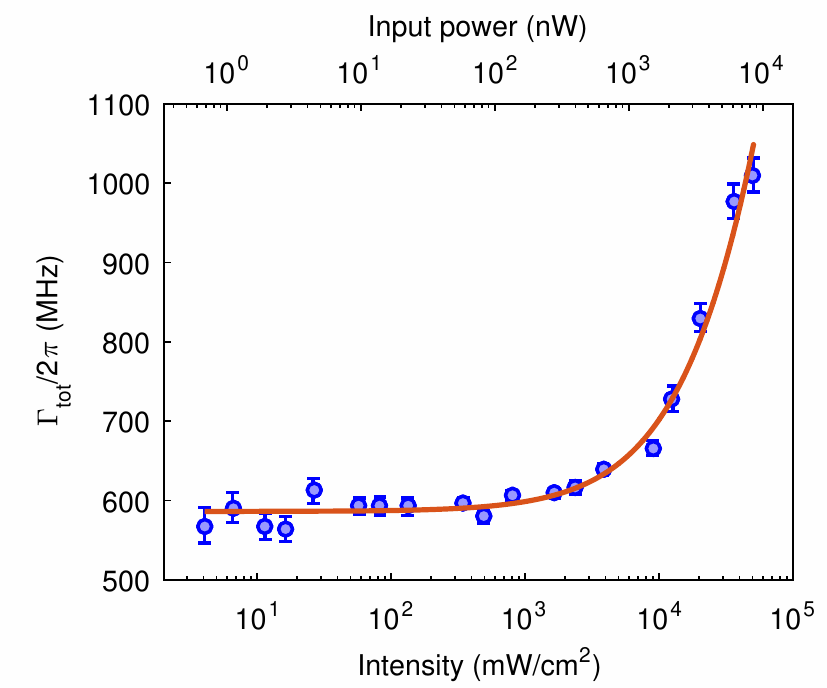}
    \caption{Saturation behavior in the ring resonator. Blue dots show the Lorentz width versus input probe power (top axis) extracted from fits to the transmission data. Error bars show 95$\%$ confidence intervals for the determined widths. The red curve is a fit of $\Gamma_{\textrm{tot}}$ with initial line width $\Gamma_0/2\pi=\SI{586}{\mega\hertz}$ and saturation power $P_{\textrm{sat}}\approx\SI{3.4}{\micro\watt}$. From $P_{\textrm{sat}}$ we estimate the corresponding mean intensity of the evanescent field (bottom axis).}
    \label{fig:fig4}
\end{figure}  

Next, we studied the saturation behavior of the atoms in the evanescent field of the ring mode. Hence we recorded a sequence of transmission spectra for different input powers with the ring resonance tuned to the atomic resonance. By fitting these spectra with our model, we extract the power dependent Lorentzian linewidth of the atoms as presented in Fig.\,\ref{fig:fig4}. The linewidth clearly exceeds the natural linewidth of rubidium already at low powers, which we attribute to transit time broadening. In order to determine the input power, at which saturation occurs, we fit the function $\Gamma_{\textrm{tot}} = \Gamma_0 (1+P_{\textrm{in}}/P_{\textrm{sat}})^{1/2}$ to the data, with initial linewidth $\Gamma_0/2\pi=\SI{586}{\mega\hertz}$ and saturation power $P_{\textrm{sat}}\approx\SI{3.4}{\micro\watt}$.
The knowledge of $P_{\textrm{sat}}$ allows us to estimate a mean intensity of the evanescent field for a given input power, by assuming $I = P_{\textrm{in}} \times I_{\textrm{sat}}/P_{\textrm{sat}}$, where $I_{\textrm{sat}}$ is the saturation intensity of the rubidium D$_{2}$ line, assuming a line width of $\Gamma_0$. By simulating the intensity distribution for our waveguide geometry we infer that $I_{\textrm{sat}}$ is reached for an atom located at the position of maximum external electric field strength with a mode power of $\sim\SI{60}{\nano\watt}$, which corresponds to a mean photon number of $\left<n\right> \approx 0.8$ photons being in the ring at any time. This number is in the same order of magnitude as the saturation (or critical) photon number as defined in cavity QED \cite{Kimble1998}, which we estimate to be $n_0 = \Gamma_{\bot}^2/2g^2 \approx 0.5$ in our case, with the transverse decay rate $\Gamma_{\bot} = \Gamma_0/2$ and the coupling parameter $g$. Together with the damping rate of the resonator $\kappa$ the corresponding cooperativity parameter for this system amounts $C = g^2/2\Gamma_{\bot} \kappa\approx 3\times 10^{-3}$, which yields a critical atom number of $N=1/C\approx 300$. 

Despite the Doppler and transit time broadening the cooperativity can be enhanced by reducing the mode volume and increasing the quality factor of the resonators. This obviously requires a different resonator design using for example photonic crystal cavities which provide extremely small mode volumes \cite{Painter2009,Scherer2001} or microresonators with ultra high $Q$ factors \cite{Aoki2006,Spillane2005}. Additionally, the photon lifetime in a resonator, and therefore the $Q$ factor, can be significantly increased by the use of slow-light effects \cite{Huet2016}. Although cooperativity values as high as in cold atom experiments are not feasible, the integration of thermal atoms with nanophotonic resonators still promises a scalable and integrable approach for nonlinear optics on the single photon level and quantum networks.
A system of coupled ring resonators also provides a platform for topological edge modes of guided light fields \cite{Hafezi2013}. Here, the interfacing with a nonlinear medium like an atomic vapor could be utilized to induce photon-photon interactions \cite{Hafezi2013_2}.

\section*{Funding Information}
We acknowledge support by the ERC under contract number 267100 and the Deutsche Forschungsgemeinschaft (DFG) with the project number LO1657/2. R.R. acknowledges funding from the Landesgraduiertenf\"orderung Baden-W\"urttemberg, H.K acknowledges support from the Carl-Zeiss-Foundation. N.G. acknowledges support by the Karlsruhe School of Optics and Photonics (KSOP).

%
%

\end{document}